\documentclass[printer]{aa}  
\usepackage{graphicx}
\usepackage{txfonts}
\usepackage{dblfloatfix}
\usepackage{natbib}
\bibpunct{(}{)}{;}{a}{}{,} 
\graphicspath{{figs/},{./}}

\newcommand{\pdv}[2]{\frac{\partial #1}{\partial #2}}
\newcommand{\bjoy}{b_{\text{joy}}}
\newcommand{\blat}{b_{\text{lat}}}
\def\ssn{S}

\def\degdot{{}^{\circ}\hspace{-0.3em}.}
\def\fsph{f_{\mbox{\scriptsize\rm sph}}}
\newcommand\ovl{\overline{\lambda_0}}
\def\devTQ{\mbox{dev}_{\scriptsize TQ}}
\def\devLQ{\mbox{dev}_{\scriptsize LQ}}

\newcommand\citen[1]{\citealt{#1}}

\usepackage{color}

\definecolor{orange}{RGB}{232,116,0}

\newcommand{\revision}{}

\begin{document} 

\title{Role of observable nonlinearities in solar cycle modulation}
   \author{M. Talafha\inst{1}
          \and
          M. Nagy\inst{1}
          \and
          A. Lemerle\inst{2,3}
          \and
          K. Petrovay\inst{1}
          }
   \institute{$^1$ELTE E\"otv\"os Lor\'and University, Institute of
   Geography and Earth Sciences, Department of Astronomy, 
              Budapest, Hungary\\
              $^2$Coll\`ege de Bois-de-Boulogne, Montr\'eal, QC, Canada\\
              $^3$D\'epartement de Physique, Universit{\'e} de
              Montr\'eal, Montr\'eal, QC, Canada\\
              \email{m.talafha@astro.elte.hu
          }
             }
   \date{Received 03 November 2021; accepted Accepted: 03 January 2022}

\titlerunning{Observable nonlinearities in solar cycle modulation}


  \abstract
   {
   Two candidate mechanisms have recently been considered with regard to the nonlinear modulation 
   of solar cycle amplitudes. Tilt quenching (TQ) comprises the negative feedback between
   the cycle amplitude and the mean tilt angle of bipolar active regions relative to
   the azimuthal direction. Latitude quenching (LQ) consists of a positive
   correlation between the cycle amplitude and average emergence latitude of active
   regions. 
   }
   {
   Here, we explore the relative importance and the determining factors
   behind the LQ and TQ effects.
   }
   {
   We systematically probed the degree of nonlinearity induced by TQ and LQ, as well as a combination of both using a grid based on surface flux transport (SFT) models. 
   The roles played by TQ and LQ are also explored in the successful 2x2D dynamo model, which has been
   optimized to reproduce the statistical behaviour of real solar cycles.
   }
{
The relative importance of LQ versus TQ is found to correlate with the ratio
$u_0/\eta$ in the SFT model grid, where $u_0$ is the meridional flow amplitude
and $\eta$ is the diffusivity. An analytical interpretation of this result is given,
further demonstrating that the main underlying parameter is the dynamo
effectivity range, $\lambda_R$, which is, in turn, determined by the ratio
of equatorial flow divergence to diffusivity. The relative importance
of LQ versus TQ is shown to scale as $C_1+C_2/\lambda_R^2$. The presence
of a latitude quenching effect is seen in the 2x2D dynamo,
contributing to the nonlinear modulation by an amount that is comparable to
TQ. For other dynamo and SFT models considered in the literature, the
contribution of LQ to the modulation covers a broad range -- from entirely insignificant to serving as a dominant source of feedback. On the other
hand, the contribution of a TQ effect (with the usually assumed
amplitude) is never shown to be negligible.
}
   {}


   \maketitle
%

\section{Introduction}
\label{sect:intro}

The origin of intercycle variations in the series of 11-year solar
activity cycles and the possibilities to  predict them have been
the focus of intense research over the past decade
(\citen{Petrovay:LRSP2}, \citen{Charbonneau:LRSP2}, \citen{Nandy:review}).
In a physical system displaying periodic or quasiperiodic behaviour,
intercycle variations must be due to either inherent nonlinearities or
external forcing. A plausible example is the way in which the strictly sinusoidal,
periodic solutions of the simple linear oscillator equation turn into
less regular quasiperiodic variations when nonlinear or stochastic
forcing terms are added. Indeed, under some simplifying assumptions,
the equations describing a solar flux transport dynamo can be
truncated to a nonlinear oscillator equation (\citen{Lopes+:oscillrev}).

Here, "external'' forcing  refers to the physical model
under consideration and not necessarily to the object of study, namely: in mean-field
models of the solar dynamo, stochastic variation triggered by
small-scale disturbances that are not explicitly modelled is the most plausible
source of such forcing. Possibilities for making solar cycle predictions are
determined by the nature and interplay of these nonlinear and stochastic
effects.

Based on its observed characteristics, the solar dynamo is likely to be an
$\alpha\omega$-dynamo, wherein strong toroidal magnetic fields are
generated by the windup of a weaker poloidal (north--south) field by
differential rotation. The strong toroidal field buried below the
surface then occasionally and locally emerges to the surface and
protrudes into the atmosphere in the form of east--west oriented
magnetic flux loops that are observed as solar active regions. Besides their
dominant azimuthal magnetic field component, these loops, presumably
under the action of Coriolis force, also develop a poloidal field
component which is typically oriented opposite to the preexisting
global poloidal field.  The evolution of the poloidal magnetic field
on the solar surface (where it is effectively radial) can be followed
on solar full-disk magnetograms and found to be well described by
phenomenological surface flux transport (SFT) models
(\citen{Jiang+:SFTreview}). In such SFT models, the evolution of the
radial magnetic field at the surface is described by an
advective-diffusive transport equation, with an added source term
representing emerging active regions and an optional sink term to
describe the effects of radial diffusion. Advection is attributed to the
poleward meridional flow, while diffusion is due to supergranular
motions.

At solar minimum, the field is found to be strongly concentrated at the
Sun's poles, so the dominant contribution to the solar dipole moment, namely:\ 
\begin{equation}                       
    D(t) = \frac32 \int_{-\pi/2}^{\pi/2} 
    {B}(\lambda,t)\sin\lambda\cos\lambda\, \mathrm{d}\lambda,
 \label{eq:dipmom}
\end{equation}
(where $\lambda$: latitude, $t$: time, $B$: azimuthally averaged 
field strength) comes from the polar caps. It is this polar field that
serves as seed for the windup into the toroidal field of the next solar
cycle, so for intercycle variations it is sufficient to consider the
azimuthally averaged SFT equation. This is expressed as:\ 
\begin{eqnarray}
\label{eq33}
 \pdv Bt &=& 
 \frac{1}{R\cos{\lambda}}\pdv{}\lambda(B\,u\,\cos{\lambda})
 \nonumber \\
 &&+\frac{\eta}{R^2\cos{\lambda}}
 \pdv{}\lambda\left(\cos{\lambda}\pdv B\lambda\right)
 -\frac{B}{\tau} + s(\lambda,t),
\label{eq:transp}
\end{eqnarray}
where $R$ is the solar radius, $\eta$ is the supergranular
diffusivity, $U$ is the meridional flow speed, $\tau$ is the time
scale of decay due to radial diffusion, and $s$ is the source
representing flux emergence.

As Eq.~(\ref{eq:transp}) is linear, the amplitude of the dipole
moment, $D_{i}$, built up by the end of cycle $i-1$ is uniquely set by
the inhomogeneous term, $s$. The windup of the poloidal field,  $D_i$,
in turn, is expected to be linearly related to the peak amplitude of
the toroidal field in cycle, $i$, and, hence, to the amplitude of the
next activity cycle --- this is indeed confirmed by both observations
and dynamo models (\citen{Munozjara+:polar.precursor},
\citen{Kumar2021}). The only step where nonlinearities and stochastic
noise can enter this process is thus via flux emergence, linking the
subsurface toroidal field to the poloidal source $s(\lambda,t)$ at the
surface. As $s(\lambda,t)$ is directly accessible to observations,
this implies that the nonlinearities involved may, in principle, be
empirically constrained.

Indeed, \cite{Dasi-Espuig+} reported that the average tilt of the axis of
bipolar active regions relative to the azimuthal direction
is anticorrelated with regard to the cycle amplitude. Starting with the pioneering work
of \cite{Cameron+:tiltprecursor}, this tilt quenching (TQ) effect
has become widely used in solar dynamo modeling, especially as the
product of the mean tilt with the cycle amplitude has been shown to be a
good predictor of the amplitude of the next cycle. The observational
evidence, however, is still inconclusive, 
and the numerical form of the effect remains insufficiently constrained 
{ (see, however, \citen{jiao2021} for more robust recent
evidence { and also \citen{Jha2020} for the effect's dependence on
field strength}). }

\cite{Jiang:nonlin} recently called attention to another nonlinear
modulation mechanism: latitude quenching (LQ). This is based on the
emprirical finding based on an analysis of a long sunspot record, which shows that
the mean latitude where active regions emerge in a given phase of the
solar cycle is correlated with the cycle amplitude. From higher latitudes, a
lower fraction of leading flux can manage become diffused across the
equator, leaving less trailing flux to contribute to the polar fields.
Therefore, the correlation found here represents a negative feedback
effect. Assuming a linear dependence for both the mean tilt and the mean
latitude on cycle amplitude, based on one particular SFT setup,
\cite{Jiang:nonlin} found that TQ and LQ yield comparable
contributions to the overall nonlinearity in the process of the
regeneration of the poloidal field from the poloidal source. The net
result showed that the net dipole moment change during a cycle tends to be
saturated for stronger cycles.

Our objective in the present work is to further explore the respective
roles played by TQ and LQ in the solar dynamo. In Section 2, we extend
the analysis to other SFT model setups and study how the respective
importance of TQ and LQ depend on model parameters. In Section 3, we
further consider a case where the form of TQ is explicitly assumed to
be nonlinear, namely: the 2x2D dynamo model. Our conclusions are summarized in
Section 4.


\section{Quadratic nonlinearities in a surface flux transport model}
\label{sect:SFT}

\subsection{Model}
\label{sect:model}

For the study of how different nonlinearities compare in SFT models
with different parameter combinations, our model setup is a
generalization of the approach of \cite{paper1} (hereafter
Paper~1).

In the prior study, our intention was to model ``typical'' or ``average''
solar cycles, so that our source function would not consist of individual
ARs but a smooth distribution (interpreted as an ensemble average)
representing the probability distribution of the emergence of
leading and trailing polarities on the solar surface.
By nature, this source is thus axially symmetric, so our whole SFT
model is reduced to one dimension, as described by
Eq.~(\ref{eq:transp}).

For the meridional flow, we consider a sinusoidal profile with a dead zone 
around the poles,
\begin{equation}
\label{eq5}
    u_{c} =
\left\{
 \begin{array}{ll}
 u_{0}\sin(\pi\lambda/\lambda_{0})  
     & \mbox{if } |\lambda| < \lambda_{0} \\
 0 & \mbox{otherwise, } 
 \end{array}
\right.  
\end{equation}
with $\lambda_0=75^\circ$. This profile has been used in many other
studies, including \cite{Jiang:nonlin}, who took the values  
$u_0=11\,$m$/$s, $\eta=250\,$km$^2/$s, and $\tau=\infty$.

Our source term is a generalization of the source used in Paper~1. The
source is a smooth distribution representing an ensemble average or
probability distribution of the emergence of leading and trailing polarities
on the solar surface (\citen{Cameron2007}, \citen{Munozjara2010}), 
represented as a pair of rings of opposite magnetic polarities:
\begin{eqnarray}
s(\lambda,t)&=&
kA_n \fsph s_1(t)
s_2\left[\lambda;\lambda_0(t)-\Delta\lambda(t),\delta\lambda\right]
\nonumber \\
&&-kA_n s_1(t)
s_2\left[\lambda;\lambda_0(t)+\Delta\lambda(t),\delta\lambda\right] 
\nonumber \\
&& +kA_n s_1(t)
s_2\left[\lambda;-\lambda_0(t)-\Delta\lambda(t),\delta\lambda\right] 
\nonumber \\
&& -kA_n \fsph s_1(t)
s_2\left[\lambda;-\lambda_0(t)+\Delta\lambda(t),\delta\lambda\right],
\label{eq:source}
\end{eqnarray}
where $k=\pm 1$ is a factor that depends on the sign of the toroidal
field and 
$A_n$ is the amplitude for cycle $n$. 
In the numerical implementation of this source profile, care was taken to 
ensure zero net flux on the spherical surface by reducing the amplitude of the
equatorward member of each pair by an appropriate sphericity factor
$\fsph$.

$s_1$ is the time profile of activity in a cycle while 
$s_2$ characterizes the latitude dependence of activity at a given cycle phase. 
In contrast to Paper~1 where a series of identical cycles was
considered, here we allow for intercycle variations, so $A_n$ will be
different for each cycle.

The time profile of solar activity in a typical cycle was determined
by \cite{hathaway1994shape} from the average of many cycles as:
\begin{equation}
    s_1(t)= at^{3}_{c}/[\exp(t^{2}_{c}/b^{2})- c],
    \label{eq:ampli}
\end{equation}
with $a = 0.00185$, $b = 48.7$, $c= 0.71$, where $t_c$ is the time since the
last cycle minimum.

The latitudinal profile
$s_2\left[\lambda;\lambda_0(t),\delta\lambda\right]$ is a Gaussian
migrating equatorward during the course of a cycle:
\begin{equation}
s_2(\lambda;\lambda_0,\delta\lambda) = \frac{\delta\lambda_0}{\delta\lambda}
\exp\left[-(\lambda-\lambda_0)^2/2\delta\lambda^2\right], 
\end{equation}
with the constant fixed as $\delta\lambda_0=6{\degdot}26$.
Following the empirical results of \cite{Jiang+:1700a}, 
the standard deviation is given by:
\begin{equation}
\label{eq:fwhm}
\delta \lambda = [ 0.14+1.05(t/P)-0.78(t/P)^{2}]\lambda_{0} 
,\end{equation}
where $P=11\,$year is the cycle period. We note that alternate fitting
formulae were determined by \citen{Lemerle1} on the basis of a
single cycle.

We obtain the latitudinal separation $2\Delta\lambda$ of the rings as a
consequence of Joy's law, while the mean latitude $\lambda_0$ was
again empirically determined by \cite{Jiang+:1700a}. These
parameters, in turn, are now dependent on cycle amplitude introducing
two nonlinearities into the problem.

{\it Tilt Quenching (TQ):} Joy's law varies depending on the amplitude
of the cycle as
\begin{equation}
\Delta\lambda=1{\degdot}5\,{\sin\lambda_0} 
  \left( 1-\bjoy\frac{A_n-A_0}{A_0}\right) .
\label{eq:mod_joyslaw}
\end{equation}
Here, $A_0$ is an arbitrary reference value corresponding to a
``typical'' or average cycle amplitude, as set by the nonlinear
limitation of the dynamo. Equation~(\ref{eq:mod_joyslaw}), thus,
measures the departure of the feedback $\Delta\lambda=f(A_n)$ from
its value in this reference state; its assumed linear form should be
valid for a restricted range in $\Delta\lambda$ and is not
incompatible with the rather lax observational constraints,--- cf.
Fig.~1b in \cite{Jiang:nonlin}.
The nonlinearity parameter $\bjoy$ is another free parameter to be
explored; our reference value for it will be 0.15.\footnote{We note that
alternate functional forms for Joy's law, admitted by the
observational constraints, such as
$\Delta\lambda\propto\lambda$ or
$\Delta\lambda\propto\sin^{1/2}\lambda$ have also been tested in the
context of the linear case (Paper 1) and found to have only a minor
impact on the results.}

{\it Latitude Quenching (LQ):} following \cite{Jiang+:1700a} for the
mean latitude $\lambda_0$ of activity at a given phase of cycle $i$ we have:
\begin{equation}
\label{eq:lambdan}
\lambda_0(t;i)  =  [ 26.4 - 34.2 (t/P) + 16.1(t/P)^2 ] 
 (\lambda_{i}/14{\degdot}6),
\end{equation}
with the mean latitude in cycle $i$ given by:
\begin{equation}
\lambda_i = 14{\degdot}6+\blat\frac{A_n-A_0}{A_0}.\\
\label{eq:latit_quen}
\end{equation}
Here, the nonlinearity parameter is fairly tightly constrained by the
empirical results of \cite{Jiang+:1700a} as $\blat\simeq 2.4$.

For the SFT model parameters, our model grid will be a subset of the
grid described in Paper~1. For $\tau,$ we only consider two values here:
a decay time scale of eight years, comparable to the cycle length and
supported by a number of studies (see Paper~1 and references therein),
and the reference case with effectively no decay ($\tau=${\revision $\infty$} years).
For $u_0=10\,$m/s, the full set of $\eta$ values in the grid is
considered; and similarly for $\eta=500\,$km$^2/$s, the full set of
$u_0$ values is studied (as listed in Table~\ref{table:dev}).
Furthermore, for the study of the effects of individual
nonlinearities, for each parameter combination in the SFT model four
cases will be considered, as listed in Table~\ref{table:cases}.
\begin{center}
        \begin{table}
                \caption{Notation of cases with different combinations
                of the nonlinearity parameters}
                \label{table:cases}
                \begin{tabular}{lcc}
                        Case & $\blat$ & $\bjoy$ \\
                        \hline
                        noQ & 0 & 0 \\
                        TQ & 0 & 0.15 \\
                        LQ & 2.4 & 0 \\
                        LQTQ & 2.4 & 0.15
                        \end{tabular}
        \end{table}
\end{center}

For each combination of the parameters $u_0$, $\eta$, $\tau$, $\blat$,
and $\bjoy$ we run the SFT code solving Eq. (\ref{eq:transp}) for
1000 solar cycles to produce a statistically meaningful sample of
cycles.

\begin{figure}[htb]
\centering
\includegraphics[width=\hsize]{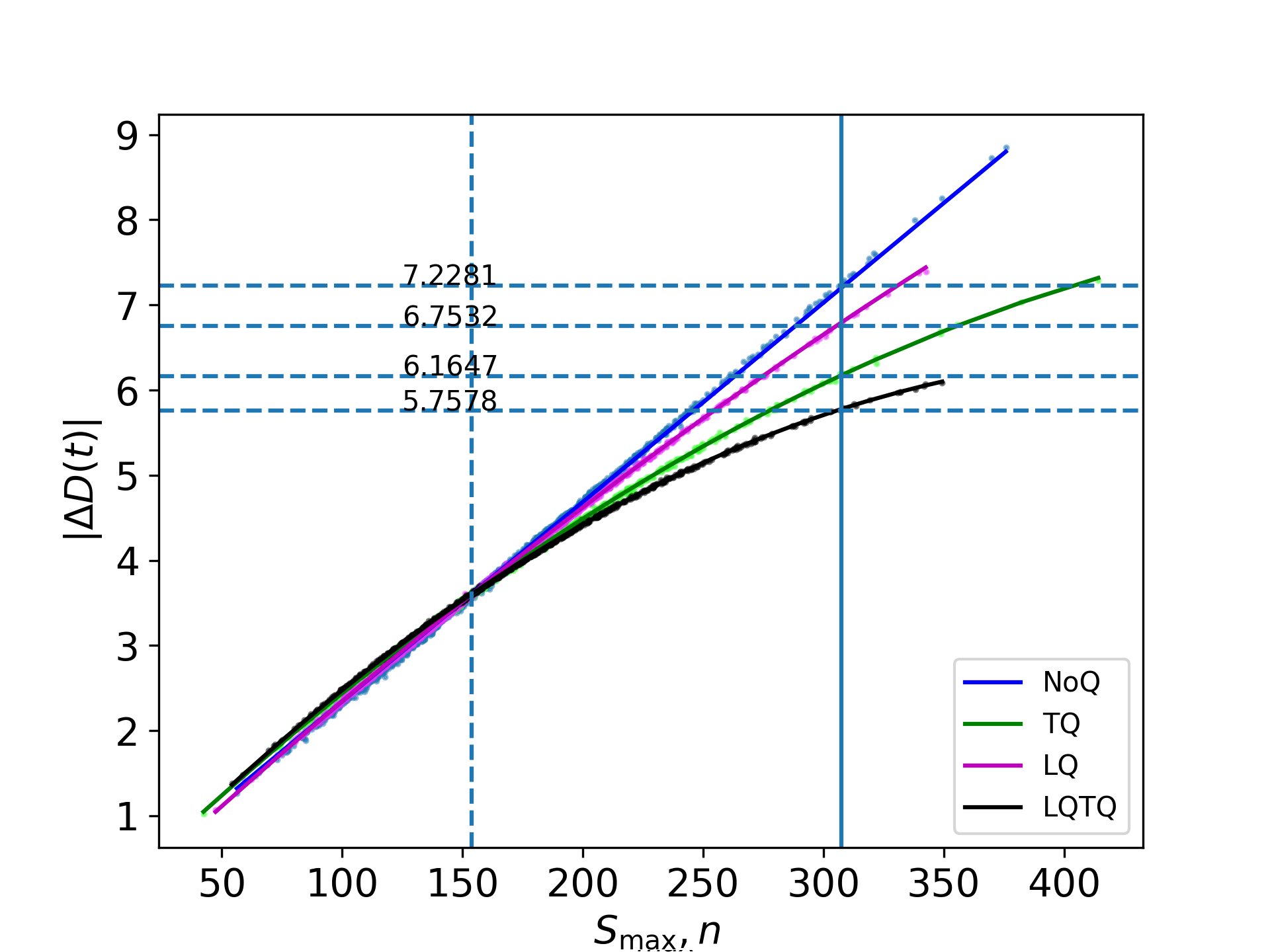}
    \caption{%
Net contribution of a cycle to the solar dipolar moment vs. cycle amplitude for
the linear case and for the cases with tilt quenching and/or latitude
quenching, with quadratic fits. 
Parameter values for this run were $u_0=5\,$m$/$s, $\eta=250\,$km$^2/$s 
and $\tau = 8$ years.
}
\label{fig:quad_fit}
\end{figure}

\begin{figure}[htb]
\centering
\includegraphics[width=\hsize]{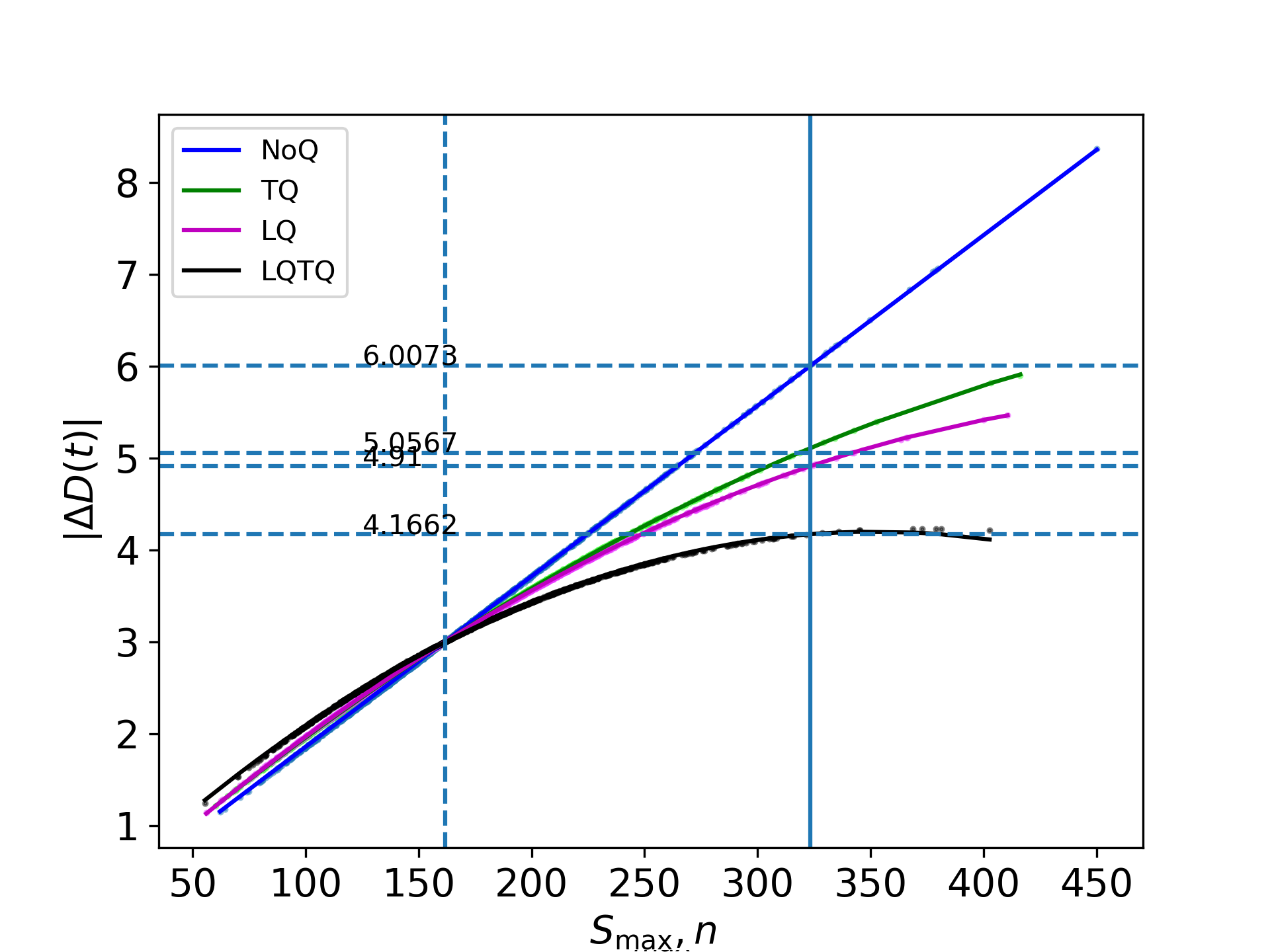}
    \caption{%
Net contribution of a cycle to the solar dipolar moment vs. cycle amplitude for
the linear case and for the cases with tilt quenching and/or latitude
quenching, with quadratic fits. 
Parameter values for this run were $u_0=20\,$m$/$s, $\eta=600\,$km$^2/$s 
and $\tau$ = $\infty$ years.}
\label{fig:dev_180}
\end{figure}

\subsection{Results}
\label{sect:SFTresults}

In order to study how the nonlinearities affect the dipole moment
built up during cycles of different amplitudes, here we consider the
purely stochastic case where individual cycle amplitudes are drawn
from a statistical distribution  without any regard for previous solar
activity. The factor $k=\pm 1$ in Eq.~(\ref{eq:source}) is assumed
to alternate between even and odd cycles. The form of the statistical
distribution chosen is not important for our current purpose but, for
concreteness, fluctuations are represented as multiplicative noise
applied to the amplitude of the poloidal field source. This noise then
follows lognormal statistics:
$
A_n=A_{0}  \times 10^{G} 
$,
where $G(0,\sigma)$ is a gaussian random variable of zero mean and
standard deviation of $\sigma = 0.13$. The coefficient for an average
cycle is set to $A_0=0.001 \,e^{7/\tau}$ to ensure that the resulting
dipolar moments roughly agree with observed values (in Gauss). 

{\revision
We note that the latitude-integrated amplitude of our source term $s(\lambda,t)$
must be linearly related to the observed value of the
smoothed international sunspot number:
\[ \ssn =K\int_{-\pi/2}^{\pi/2} s(\lambda,t)\,d\lambda  . \] 
The proportionality coefficient, $K,$ was arbitrarily tuned to 
closely reproduce the observed mean value of $S$ for each value of
$\tau$.
}

Introducing the notation $r= e^{-11/\tau}$ (with $\tau$ in years), the net
contribution of cycle $n$ to the change in dipolar moment is given by
$\Delta D_n = D_{n+1}-r D_n$.
Figures~\ref{fig:quad_fit} and \ref{fig:dev_180} display $|\Delta D|$
against the cycle amplitude $S_n$ for some example runs.

The linear case ({\revision noQ}) is obviously well suited to a linear fit, while
all three nonlinear cases are seen to be well represented by quadratic
fits forced to intersect the linear fit at the mean value of $S_n$. 

\begin{figure}[htb]
\centering
\includegraphics[width=\hsize]{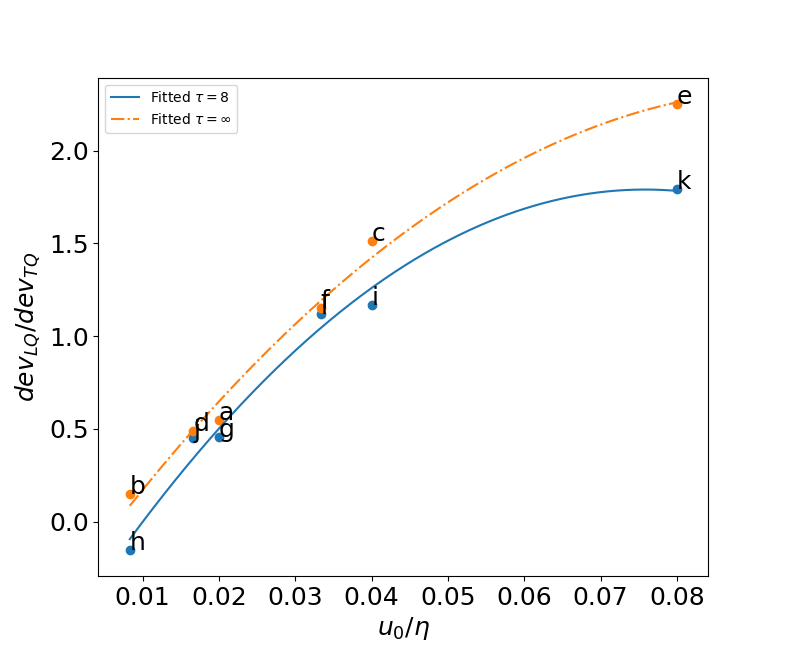}
    \caption{%
Relative importance of LQ vs TQ, plotted against the
parameter $u_0/\eta$ for the parameter combinations listed in 
Table~\ref{table:dev}. Separate quadratic fits for the case
$\tau$ ={\revision $\infty$} (orange) and $\tau=8$ years (blue) are shown.
}
\label{fig:ratio_u0_eta_all}
\end{figure}

The respective importance of TQ and LQ is seen to differ for these
cases. For a quantitative characterization of the influence of
nonlinearities we use the deviation of $|\Delta D_n|$ from the linear
case at an arbitrary selected value of $S_n$ (roughly twice the mean),
as illustrated by the solid vertical lines in
Figs.~\ref{fig:quad_fit}--\ref{fig:dev_180}. To give an example, in
the TQ case, the deviation is $\devTQ=7.2281-6.1647=1.0634$ for the
model in Fig.~\ref{fig:quad_fit}. The relative importance of latitude
quenching versus tilt quenching is then characterized  by the ratio of
these deviations for the LQ and TQ cases, $\devLQ/\devTQ$. Comparing
these values to several parameter combinations reveals that the
relative importance of the two effects is primarily determined by the
ratio $u_0/\eta$. This is borne out in Fig.~\ref{fig:ratio_u0_eta_all}.

\begin{table}
\begin{center}
\caption{Parameters of the models and their dynamo effectivity range 
}
\label{table:dev}
\begin{tabular}{crrrr}    
\hline  \\  
case &  $\tau$ [yr] &$u_0$ [m$/$s] & $\eta$ [km$^2/$s] & $\lambda_R$ [$^\circ$] \\
\\
\hline
a & {\revision $\infty$} &  5 & 250 & 12.35 \\
b & {\revision $\infty$} &  5 & 600 & 18.52 \\
c & {\revision $\infty$} & 10 & 250 &  9.44 \\
d & {\revision $\infty$} & 10 & 600 & 13.41 \\
e & {\revision $\infty$} & 20 & 250 &  7.79 \\
f & {\revision $\infty$} & 20 & 600 & 10.05 \\
g &   8 &  5 & 250 & 12.35 \\
h &   8 &  5 & 600 & 18.52 \\
i &   8 & 10 & 250 &  9.44 \\
j &   8 & 10 & 600 & 13.41 \\
k &   8 & 20 & 250 &  7.79 \\
l &   8 & 20 & 600 & 10.05 \\
\hline
\end{tabular}
\end{center}
\end{table}

\begin{figure}[htb]
        \centering
        \includegraphics[width=\hsize]{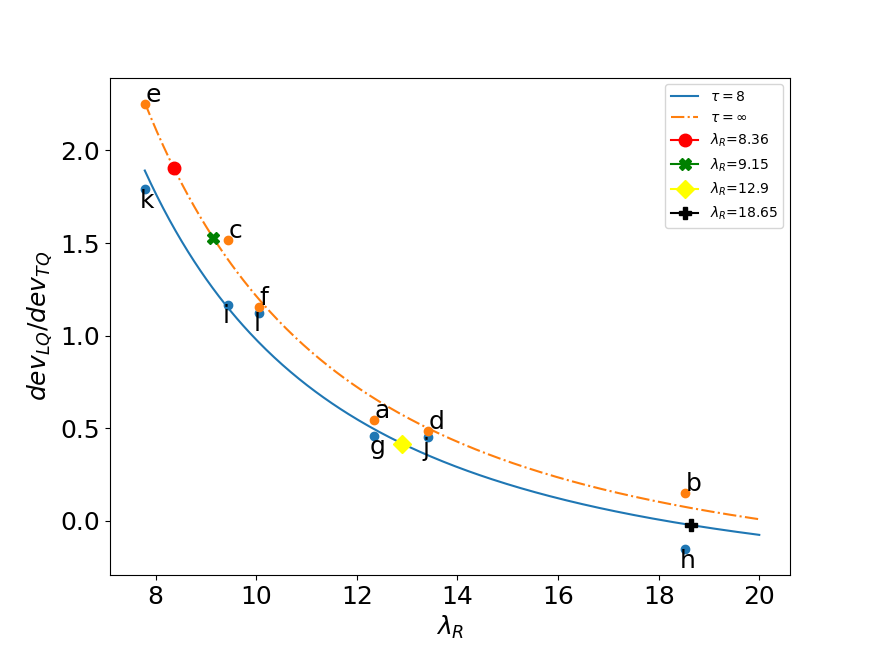}
        \caption{%
Relative importance of LQ vs TQ, plotted against the dynamo 
effectivity range, $\lambda_R,$ for the parameter combinations listed 
in Table~\ref{table:dev}. Separate fits of the form
$C_1+C_2/\lambda_R^2$ are shown for the case
$\tau =$ {\revision $\infty$} (orange) and $\tau=8$ years (blue).
Large coloured symbols mark the loci of SFT and dynamo models studied
in recent literature: 
\cite{Bhowmik_Nandy_2018} (\textcolor{red}{$\bullet$}),
\cite{Jiang+Cao} (\textcolor{green}{X}), 
\cite{Lemerle2} (\textcolor{yellow}{$\star$}),
\cite{Whitbread+:howmany} ($+$).
}
\label{fig:lambdar_quad_sep}
\end{figure}

\subsection{Interpretation}

Over a short interval of time $[t,t+dt],$ the smooth source distribution assumed,
Eq.~(\ref{eq:source}) is reduced to a bipole with a flux of
\begin{eqnarray}
\Phi(t)\,dt &=&
   2\pi A_n s_1(t) \cos\lambda_0 \,R^2 
   \int_{-\pi/2}^{\pi/2} s_2\,d\lambda \,dt \nonumber \\
   &\simeq & (2\pi)^{3/2}  A_n s_1(t) \cos\lambda_0 \,R^2 \,dt
,\end{eqnarray}
in each polarity.

The ultimate contribution of such a bipole to the solar axial dipole moment
after several years has been analytically derived by
\cite{Petrovay+:algebraic1} as: 
\begin{equation}
\delta D_U = f_\infty\, \delta D_1 \, e^{(t-t_{n+1})/\tau}
\label{eq:dDU}
,\end{equation}
where $\delta D_1$ is the initial contribution of the bipole to the global
dipole moment, $t_{n+1}$ is the time of the minimum of the next cycle, while the
asymptotic dipole contribution factor $f_\infty$ is given by:
\begin{equation}
f_\infty= \frac a{\lambda_R} \exp
\left(\frac{-\lambda_0^2}{2\lambda_R^2}\right).
\label{eq:finfty}
\end{equation}
Here, $a$ is a constant and defining $\Delta_u$ as the divergence of
the meridional flow at the equator and the dynamo effectivity range
$\lambda_R$ is a universal function of $\Delta_u/\eta$ for plausible
choices of the meridional flow profile.

For the source considered here: 
\begin{equation}
\delta D_1(t)=\frac{3\Phi}{\pi R^2}\,\Delta\lambda\,\cos\lambda_0  =
  6(2\pi)^{1/2} A_n\, \Delta\lambda\,  s_1 \, {
  \revision \cos^2\lambda_0} 
\label{eq:dD1}
\end{equation}
Combining the last three equations and integrating for the full duration of 
the cycle yields the net change in the solar dipole moment:
\begin{equation}
\Delta D_n \equiv D_{n+1}-rD_n = \int_{t_n}^{t_{n+1}} \delta D_U(t) \,dt = 
  \frac{6a}{\lambda_R} (2\pi)^{1/2} A_n I_1
\label{eq:interpret}
,\end{equation}
with
\begin{equation}
 I_1 = \int_{t_n}^{t_{n+1}} \Delta\lambda\, \cos^2\lambda_0\, s_1(t)\,
 \exp\left( \frac{t-t_{n+1}}\tau - \frac{\lambda_0^2}{2\lambda_R^2} \right) \,dt
.\end{equation}

This shows that $|\Delta D_n|$ and, hence, the nonlinear effects in
it, only depend on model parameters through $\tau$ and the dynamo
effectivity range $\lambda_R$. For a fixed meridional flow profile, as
used in this work, $\Delta_u\sim u_0$, hence $\lambda_R$ is uniquely
related to $u_0/\eta$, explaining the dependence on the latter
variable, as shown in Fig.~\ref{fig:ratio_u0_eta_all}. On the other
hand, usinq Eqs. (26) and (27) of \cite{Petrovay+:algebraic1},
$\lambda_R$ can be derived for each of our models, allowing us to plot
the underlying relation of the nonlinearity ratios with $\lambda_R$.
This plot is displayed in Fig.~\ref{fig:lambdar_quad_sep}.

In order to understand the form of this relation we recall from 
Eqs. (\ref{eq:mod_joyslaw})--(\ref{eq:latit_quen}) that
\begin{equation} \Delta\lambda \sim 1-\bjoy\frac{A_n-A_0}{A_0}  \qquad
\mbox{and} \qquad \lambda_0 \sim  1+\blat\frac{A_n-A_0}{A_0}
.\end{equation} 
Upon substituting the Eqs.{  
(\ref{eq:mod_joyslaw})--(\ref{eq:latit_quen})} into 
(\ref{eq:interpret}), the deviations from the linear case correspond to 
the terms containing $\bjoy$ and $\blat$. We will consider these terms
in the limit $\tau\rightarrow\infty$ and, relying on the mean value
theorem for integrals, time integrals will be substituted by the
product of the cycle period and the value of the integrand for an
appropriately defined average. These weighted averages are not
expected to differ from the unweighted average  $\overline{\lambda_0}$
by more than a few degrees, so they will be substituted here simply by
$\overline{\lambda_0}$ for an approximation.

In the TQ case, {\revision using Eq. (\ref{eq:mod_joyslaw}),} 
the term containing $\bjoy$ is immediately found to  scale as
\begin{equation}
\mbox{dev}_{TQ}\sim \sin\ovl \cos^2\ovl 
  \exp\left( -\frac{\ovl^2}{2\lambda_R^2}\right).
\label{eq:devtq}
\end{equation}

The LQ case is slightly more complex. In the limit
$\tau\rightarrow\infty$, the dependence on the latitude of activity
enters Eq. (\ref{eq:interpret}) through the combination 
$h(\lambda_0)=\sin\lambda_0\cos^2\lambda_0\exp (-
{\lambda_0^2}/{2\lambda_R^2})$. A Taylor expansion yields:
\begin{eqnarray}
&& h(\lambda_0+\varepsilon\lambda_0)= h(\lambda_0)
  +\varepsilon{\lambda_0} \cos\lambda_0
 \exp\left(-\frac{\lambda_0^2}{2\lambda_R^2} \right) \times \nonumber \\
&& \qquad (\cos^2\lambda_0 -2\sin^2\lambda_0 +
 \lambda_0\sin\lambda_0\cos\lambda_0 /\lambda_R^2). 
\end{eqnarray}

Substituting here $\varepsilon=\blat\frac{A_n-A_0}{A_0}$ and plugging the
expression back into Eq. (\ref{eq:interpret}), the term involving
$\blat$ is found to scale as: 
\begin{eqnarray}
&& \mbox{dev}_{LQ}\sim \ovl \cos\ovl 
  \exp\left( -\frac{\ovl^2}{2\lambda_R^2}\right) \times \nonumber \\
 && \qquad (\cos^2\ovl -2\sin^2\ovl+\ovl\sin\ovl\cos\ovl/\lambda_R^2).
\label{eq:devlq}
\end{eqnarray}

Dividing Eq. (\ref{eq:devlq}) by  (\ref{eq:devtq}) yields
\begin{equation}
\mbox{dev}_{LQ}/\mbox{dev}_{TQ}\sim
C_1(\ovl) + C_2(\ovl)/\lambda_R^2 .
\label{eq:devlqtq}
\end{equation}
This expectation is indeed confirmed by the fits in 
Fig.~\ref{fig:lambdar_quad_sep}.

\subsection{Discussion}

It is to be noted in Fig.~\ref{fig:lambdar_quad_sep} that for high
values of $\lambda_R$, the value of $\devLQ/\devTQ$ becomes slightly negative.
This is due to the circumstance that latitudinal quenching can affect
$\Delta D$ in two ways:\\
(i) by modulating the fraction of active regions above $\lambda_R$
that do not contribute to $\Delta D$, which results in a negative
feedback; (ii) by modulating the mean tilt angle of active regions in accordance
with Joy's law, which is a positive feedback effect.

For $\lambda_R\la\ovl$ the first effect dominates. However, 
when $\lambda_R$ significantly exceeds $\ovl$ the fraction of ARs 
above $\lambda_R$ becomes negligible, so only the second, positive
feedback effect remains. As a result, $\devLQ$ changes sign.

A slight dependence on $\tau$ is also noticed. This can be attributed
to the fact that for shorter values of $\tau$ the dipole moment
contribution of ARs appearing early in the cycle will have more time
to decay until the next minimum, hence, these ARs will have a lower
relative contribution to $|\Delta D|$. But the ARs in the early part
of the cycle are the ones at the highest latitudes, that is, they are most affected
by LQ via effect (i) above. Hence, for shorter values of $\tau,$
the relative importance of LQ is expected to be slightly suppressed. This
suppression may be expected to vanish towards higher values of
$\lambda_R$, in agreement with Fig.~\ref{fig:lambdar_quad_sep}.

The larger, coloured symbols in Fig.~\ref{fig:lambdar_quad_sep} mark
the positions of various dynamo and SFT models discussed in recent
publications. It is apparent that the case where LQ results in a
positive feedback is not strictly of academic interest: at least one SFT
model discussed recently, namely, that of \cite{Whitbread+:howmany} is in
this parameter range. The parameter choice for this model was
motivated by an overall optimization of observed versus simulated
magnetic supersynoptic maps (\citen{Whitbread+:SFT}).
At the other end of the scale, the low diffusivity, medium flow speed
models of \cite{Bhowmik_Nandy_2018} and \cite{Jiang+Cao} are in a
range where LQ is manifest as a negative feedback giving the
dominating contribution to the nonlinearity, although the effect of TQ
remains non-negligible.

Conversely, the locus of the SFT component of the 2x2D model of
\cite{Lemerle2} in Fig.~\ref{fig:lambdar_quad_sep} indicates that if
the TQ and LQ effects were formulated as in our model, the
contributions of TQ should dominate but the negative feedback due to
LQ would also contribute quite significantly to the nonlinearity. We take a closer look 
at this model in the following section.

{\revision

As a caveat regarding the validity of these conclusions, we recall that
the results presented in this section are based on the assumption of
one particular form of the time profile of the source function,
Eq. (\ref{eq:ampli}), with constant parameters and a strictly
fixed periodicity of 11 years. As long as deviations from it are
stochastic, this source may still be considered representative as an
ensemble average. However, there have also been indications of systematic
deviations. 

It is well known that the length of solar cycles is inversely related
to their amplitude: this effect implies an inverse correlation of the
parameter $b$ in Eq. (\ref{eq:ampli}) with $a$ (or, equivalently,
$S_{\rm max}$). It is, however, straightforward to see that a variation
in $b$ will not affect our results. If $c$ is kept  constant, as
suggested by \cite{hathaway1994shape}, $b$ is the only time scale
appearing in $s_1$, so, with $S_{\rm max}$ fixed, its variation will
simply manifest as a self-similar temporal expansion or contraction of
the profile (\ref{eq:ampli}). As a result, cycle lengths will then
vary and so will the area below the curve, namely, the total amount of
emerging flux, and, hence, the amplitude of the dipole moment source. As
a result, the dipole moment built up by the end of the cycle will 
linearly scale with $b/{\overline b}$. This scaling factor, however,
will be the same for the TQ, LQ, and subsequent cases, so the ratio between their
deviations, plotted in Figure~\ref{fig:lambdar_quad_sep}, remains unaffected.

Furthermore, \cite{jiang2018predictability} noted that actual cycle
profiles tend to show a systematic upward deviation from the profile 
(\ref{eq:ampli}) in their late phases. Such an effect would indeed
had an impact on our results somewhat. As in these late phases the activity is
concentrated near the equator, at latitudes well below $\lambda_R$,
the extra contribution to the dipole moment due to this excess is not
expected to be subject to LQ, while it is still affected by TQ. Hence,
$\devLQ/\devTQ$, plotted in Figure~\ref{fig:lambdar_quad_sep}, is
expected to be somewhat lower with this correction. This relative
difference, in turn, may be expected to vanish towards higher values of
$\lambda_R$, where LQ becomes insignificant anyway.

}


\begin{figure}[htb]
\centering
\includegraphics[width=\hsize]{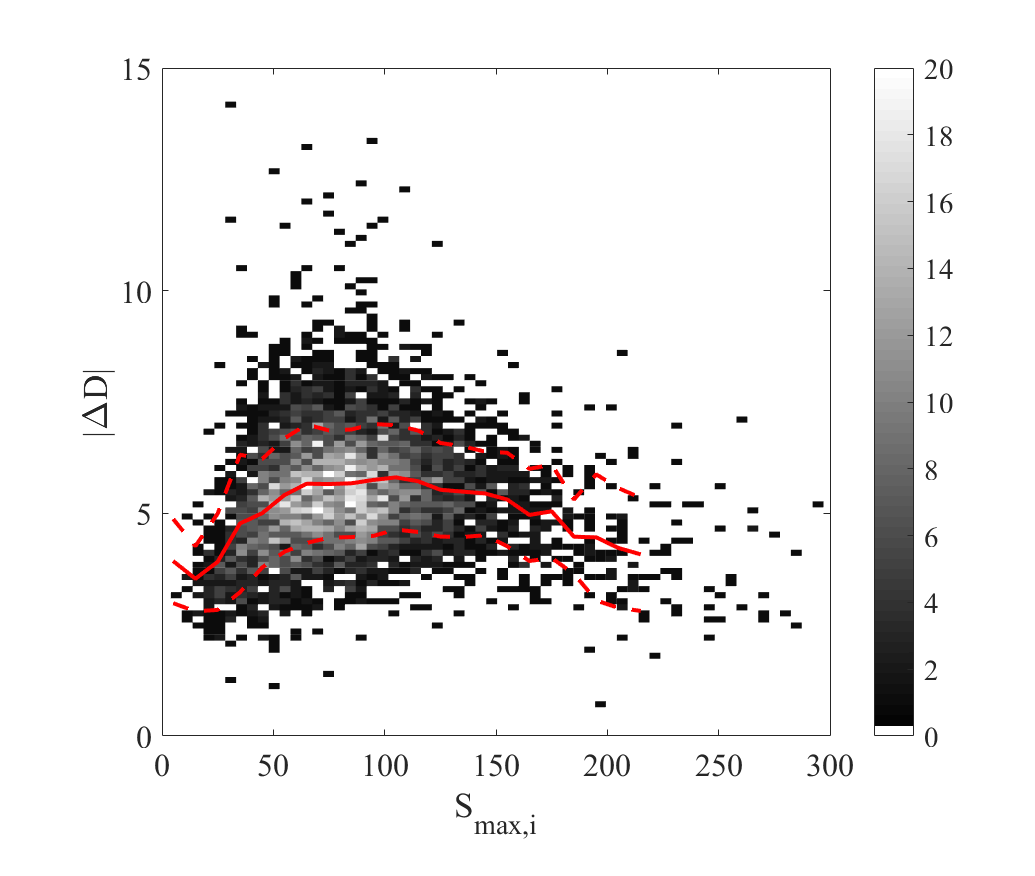}
    \caption{%
\ Contribution of all active regions in a cycle to the net change in
the solar dipole moment during dynamo cycles vs. the cycle amplitude in
the 2x2D dynamo model. 
Solid line: Mean; dashed  curves: $\pm 1\sigma$ range.
}
\label{fig:2x2D_logSSN_DeltaD}
\end{figure}

\begin{figure}[htb]
\centering
\includegraphics[width=\hsize]{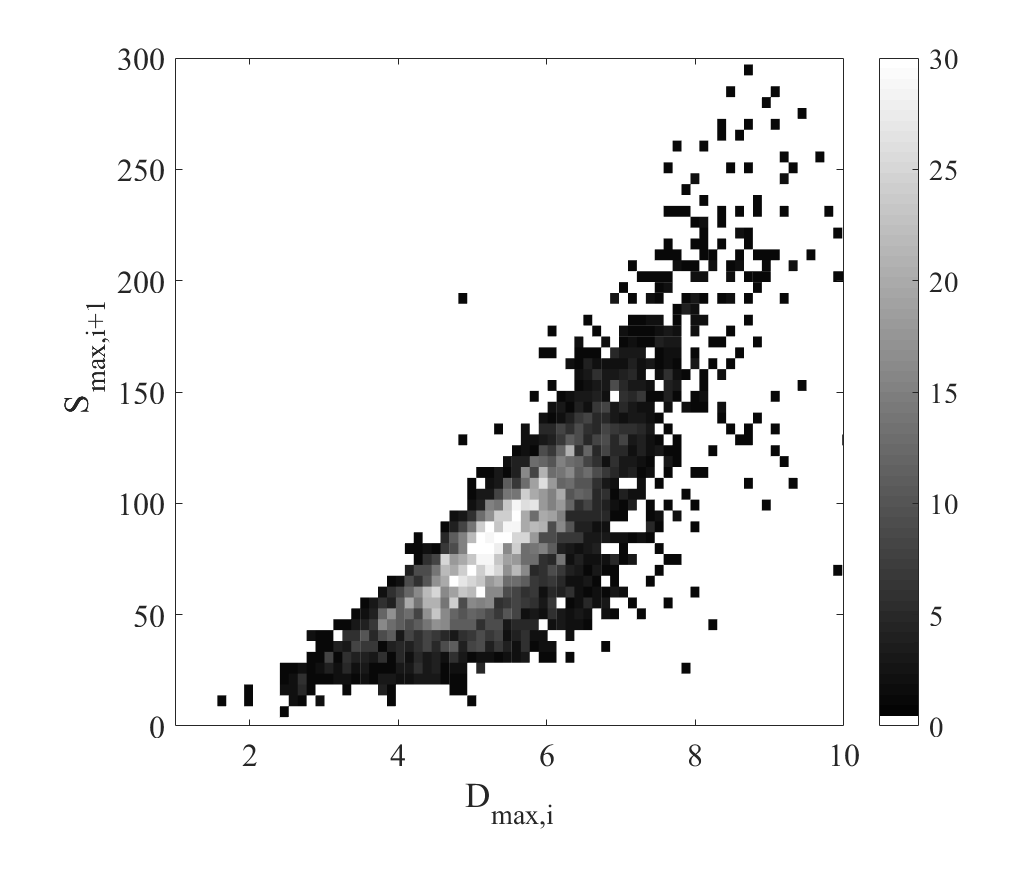}
    \caption{%
Solar dipole moment at cycle minimum vs. the amplitude of the subsequent cycle
in the 2x2D dynamo model.
}
\label{fig:2x2D_SSN-D}
\end{figure}

\begin{figure}[htb]
\centering
\includegraphics[width=\hsize]{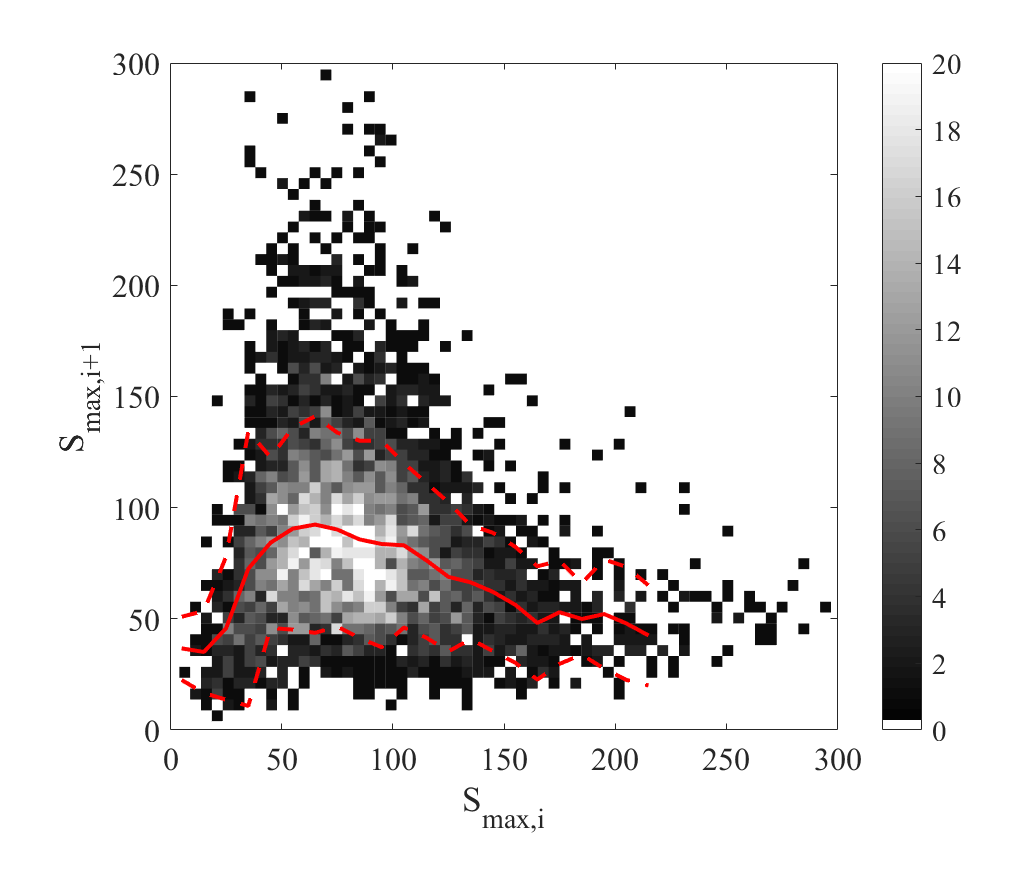}
    \caption{%
Cycle amplitude vs. the amplitude of the subsequent cycle 
in the 2x2D dynamo model.
Solid line: Mean; dashed  curves: $\pm 1\sigma$ range.
}
\label{fig:2x2D_mapping}
\end{figure}

\section{Nonlinear effects in the 2x2D dynamo model}
\label{sect:2x2D}

In this section, we consider the role of nonlinearities in a full
dynamo model,   namely, the 2x2D dynamo model (\citen{Lemerle1} and
\citen{Lemerle2}). This model, also used for solar cycle forecast
(\citen{Labonville2019}), couples a 2D surface SFT model with an
axisymmetric model of the dynamo operating in the convective zone. The
azimuthal average of the poloidal magnetic field resulting from the
SFT component is used as upper boundary condition in the subsurface
component, while the source term in the SFT model is a set of bipolar
magnetic regions (BMRs) introduced instantaneously with a probability
related to the amount of toroidal flux in a layer near the base of the
convective zone, as a simplified representation of the flux emergence
process. The model has a large number of parameters optimized for best
reproduction of the observed characteristics of solar cycle 21.

\subsection{Nonlinearity vs. stochastic scatter in the 2x2D dynamo}

Figure~\ref{fig:2x2D_logSSN_DeltaD} presents the equivalent of 
Figs.~\ref{fig:quad_fit} and \ref{fig:dev_180} for 5091 cycles
simulated in the 2x2D model. A marked nonlinearity is clearly seen:
the net contribution of a cycle to the dipole moment increases slower
than in the linear case with cycle amplitude, representing a negative feedback,
similarly to the case of the pure SFT model.

In the dynamo model, the dipole moment built up during a cycle feeds
back into the cycle as the amplitude of the toroidal field generated
by the windup of the poloidal fields (and, hence, the amplitude of the
upcoming cycle) increases with the amplitude of the poloidal field.
This is borne out in Fig.~\ref{fig:2x2D_SSN-D}. We note that the relation
is not exactly linear and there is a slight threshold effect at work, which is
presumably related to the requirement that the toroidal field must
exceed a minimal field strength for emergence.

The combined result of the two feedbacks presented in 
Figs.~\ref{fig:2x2D_logSSN_DeltaD} and \ref{fig:2x2D_SSN-D} is a
nonlinear coupling between the amplitudes of subsequent cycles as
shown in Fig.~\ref{fig:2x2D_mapping}. The convex shape of the median
curve implies that on average, very weak cycles will be followed by
stronger cycles and very strong cycles will be followed by weak
cycles. However, there is a very large amount of scatter present in
the plot. A comparison with the scatter in
Figs.~\ref{fig:2x2D_logSSN_DeltaD} and \ref{fig:2x2D_SSN-D} shows that
the main contribution to the scatter comes from
Fig.~\ref{fig:2x2D_logSSN_DeltaD}. As $\Delta D$ is uniquely
determined by the source term in the SFT equation representing flux
emergence, the main cause of random scatter around the mean nonlinear
relation is identified as the stochastic nature of the flux emergence
process, namely, the scatter in the time, location, and properties of
individual active regions (BMRs in the 2x2D model) statistically
following the evolution of the underlying toroidal field.

\begin{figure}[htb]
\centering
\includegraphics[width=\hsize]{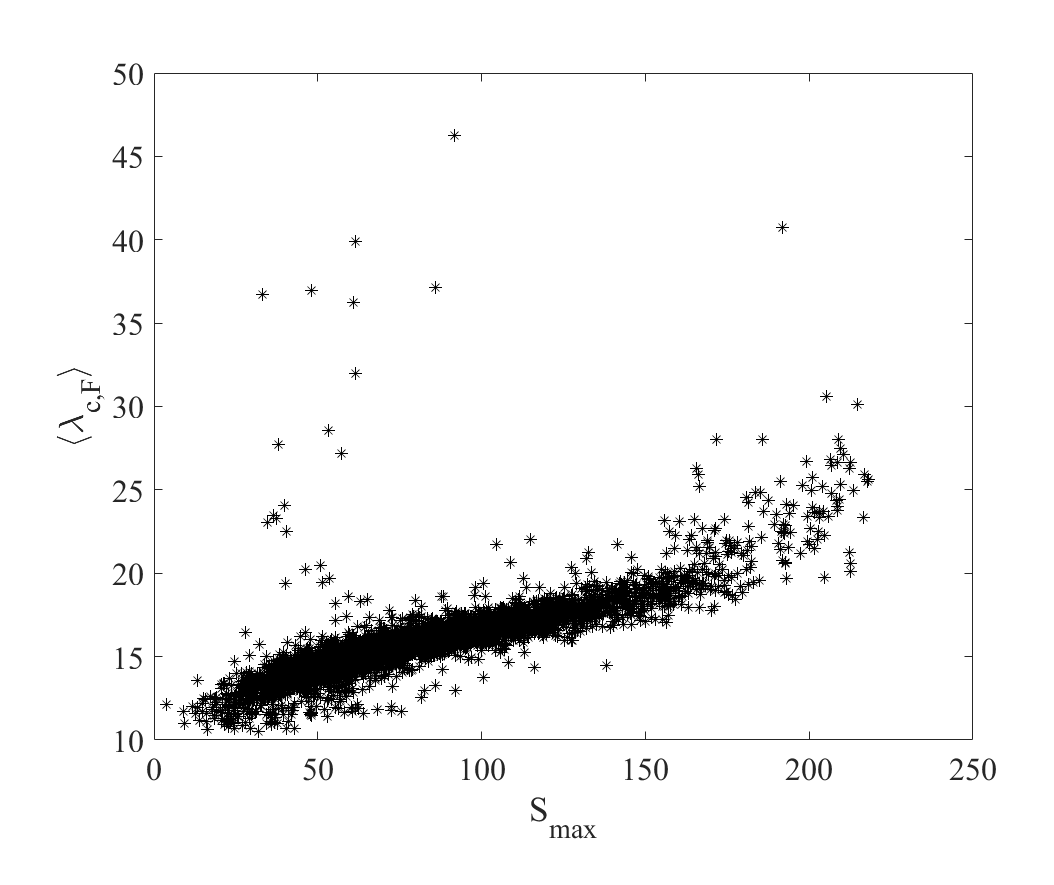}
    \caption{%
Latitude quenching in the the 2x2D dynamo model. Mean latitude of BMRs in a
cycle vs. cycle amplitude. 
}
\label{fig:LQ_2x2D_total_hemisph}
\end{figure}

\begin{figure}[htb]
\centering
\includegraphics[width=\hsize]{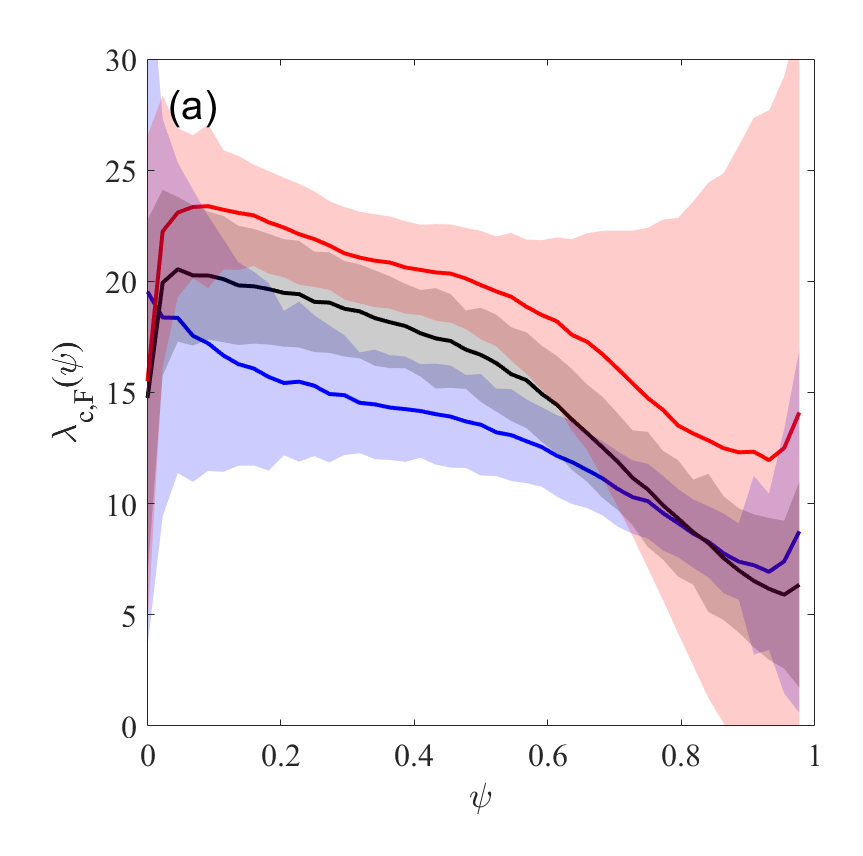}
\includegraphics[width=\hsize]{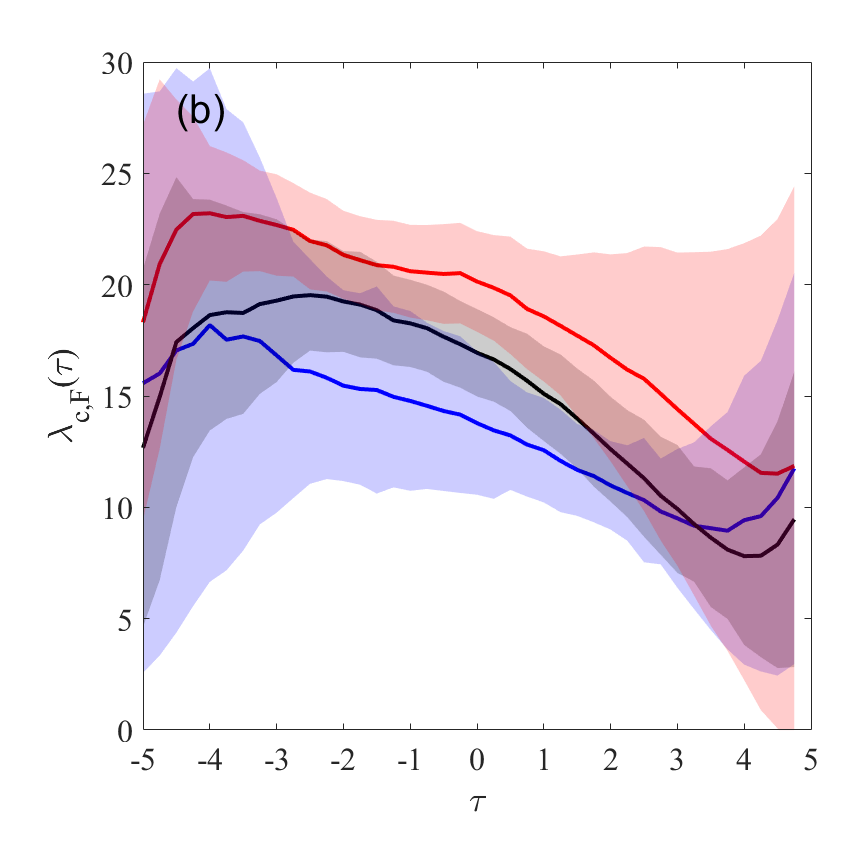}
    \caption{%
Butterfly tracks in the 2x2D dynamo model sorted by cycle amplitude: 
(a) tracks of the active latitude vs.\ cycle phase; 
(b) tracks of the active latitude vs.\ time; 
Red, black, and blue curves represent flux-weighted mean latitudes for subsamples of 
strong, average, weak cycles, respectively, as described in the text. 
Shaded areas mark the standard deviations of the actual data around
the mean curves.
}
\label{fig:meanLatitude_fluxweight_cycledep220}
\end{figure}

\subsection{Latitude quenching in the 2x2D dynamo model}

The question arises as to whether the nonlinearity plotted in
Fig.~\ref{fig:2x2D_logSSN_DeltaD} is a consequence of TQ, LQ, or both.
Apart from a threshold effect which was not found to be strong or even
necessarily needed in the optimized model, the only nonlinearity
explicitly built into the 2x2D model is TQ. The tilt angle (i.e.,
initial dipole contribution $\delta D_1$)  of a BMR is assumed to
scale with 
\begin{equation}
    \alpha_q = \frac{\alpha}{1 + (B_{\phi}/B_q)^2}
\label{eq:tiltquenching}
,\end{equation}
where $B_q$ is the quenching field amplitude.

As the latitudinal distribution of BMRs is fed into the SFT component
of the 2x2D model from the mean-field component, LQ cannot be imposed.
Nevertheless, this does not mean that LQ is not present. Indeed,
as demonstrated in Fig.~\ref{fig:LQ_2x2D_total_hemisph}, the 2x2D
model displays a very marked latitude quenching effect, with a tight
correlation between the mean emergence latitude of BMRs and cycle
amplitude. This LQ is even stronger than witnessed in the Sun, $\ovl$
varying by $\sim 10^\circ$ as $S_n/\overline{S_n}$ varies between
$0.1$ and $2$. In contrast, according to Eq.
(\ref{eq:latit_quen}), $\ovl$ only varies by about $5^\circ$  (from
$12^\circ$ to $17^\circ$) for the same range in $S_n$.  

The origin of this LQ effect is elucidated in 
Fig.~\ref{fig:meanLatitude_fluxweight_cycledep220}. Three subsamples
of all simulated cycles are considered: cycles stronger or weaker than
the mean amplitude by at least 1 standard deviation are displayed in
red or blue, respectively, while black marks cycles not farther from the
mean amplitude than 0.1 standard deviation. The plots clearly
demonstrate that tracks of latitudinal migration of these three
subsets are systematically shifted relative to each other. The
alternative possibility that all cycles follow the same track with LQ
resulting from the different temporal variation of the activity level
can thus be discarded.

\begin{figure}[htb]
\centering
\includegraphics[width=\hsize]{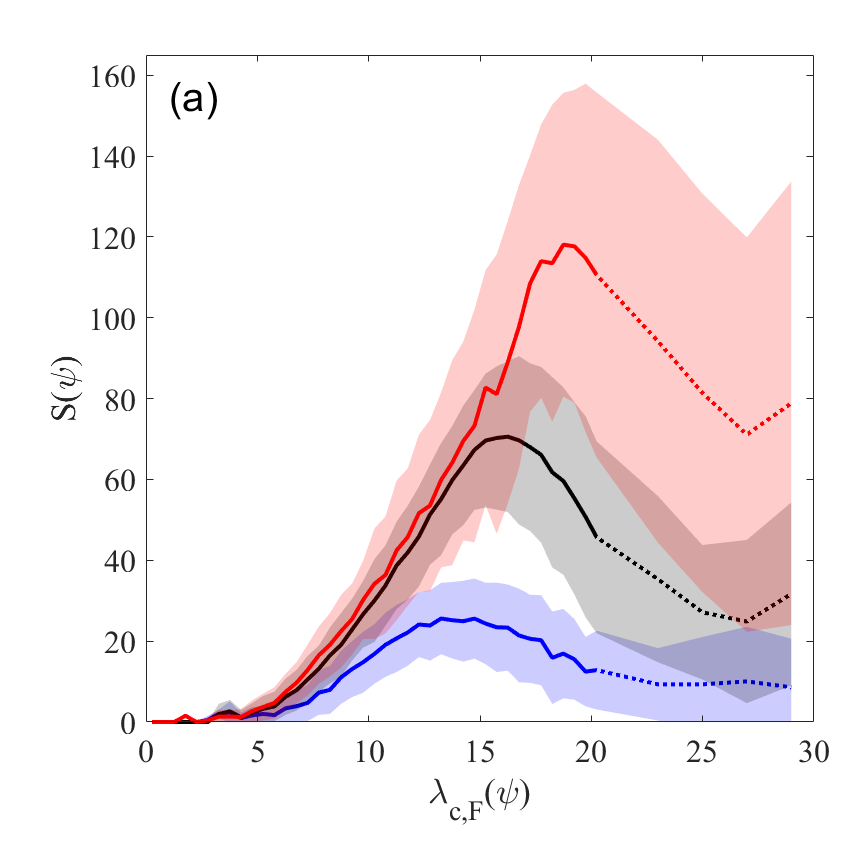}
\includegraphics[width=\hsize]{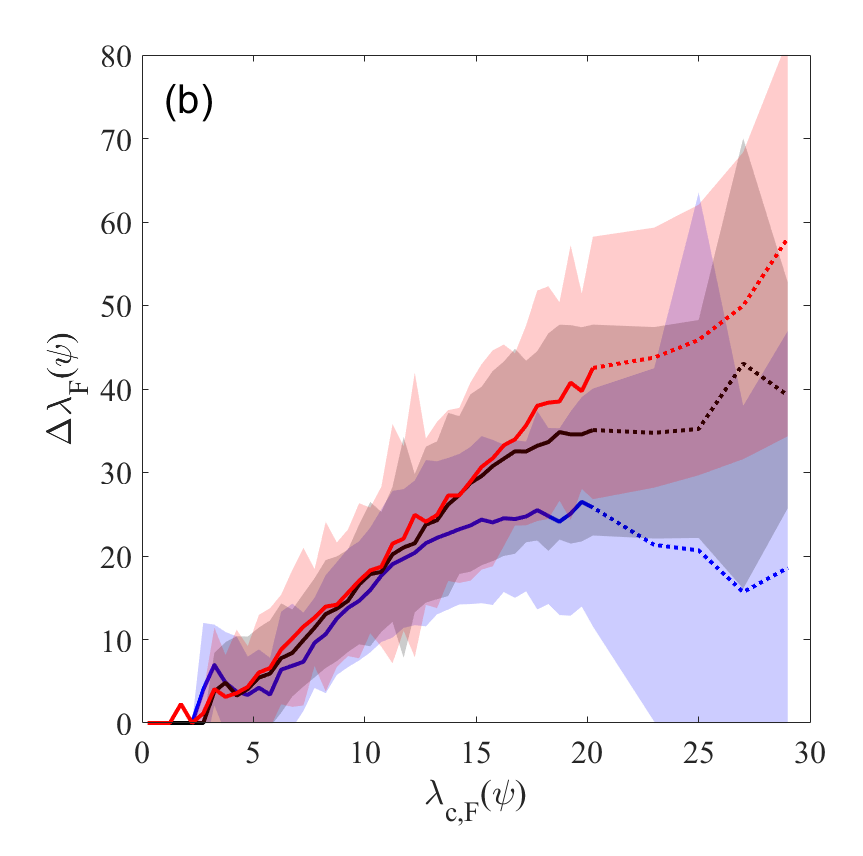}
    \caption{%
Universality of the cycle decay phase in the 2x2D dynamo model: (a) Pseudo sunspot
number against flux-weighted mean latitude of BMRs. Curves show
flux-weighted mean values; (b) Width $2\delta\lambda$ of the
latitudinal distribution of BMRs vs. their flux-weighted mean  
latitude.  Colour coding is the same as in 
Fig.~\ref{fig:meanLatitude_fluxweight_cycledep220} Shaded areas mark
the standard deviations of actual data around the mean curves. Solid
lines correspond to a $0.5^{\circ}$ binning,  dotted lines do to a
$2^{\circ}$ binning. 
}
\label{fig:universal}
\end{figure}

\cite{Cameron+Schussler:turbdiff} considered the time evolution of
mean latitude and width of the activity belts in observed solar
cycles, interpreting the decay phase of cycles as a result of the
cancellation of oppositely oriented toroidal flux bundles across the
equator by turbulent diffusion. They pointed out that this leads to a
certain universality of the characteristics of the decay phases in
different solar cycles. Our finding that the tracks of migrating
toroidal flux bundles systematically differ for strong and weak cycles
may at first sight seem to contradict this interpretation. However, a
closer look at the properties of decay phases of cycles simulated in
the 2x2D model confirms the universality posited by
\cite{Cameron+Schussler:turbdiff}. In Fig.~\ref{fig:universal}a, we
present the evolution of the (pseudo) sunspot number simulated in the
models against the latitude of activity belts, rather than against
time, while Fig.~\ref{fig:universal}b presents the evolution of the
width of the latitudinal distribution of BMRs in a similar way (see
Figs. 3 and 4 in \citen{Cameron+Schussler:turbdiff} for the
observational equivalents of these plots.) It is apparent that the
curves for the strong, average and weak subsamples are nicely aligned
towards their low-latitude (i.e., late-phase) ends, confirming the
universality proposition.

\section{Conclusion}
\label{sect:concl}

In this work, we explore the relative importance and the
determining factors behind the LQ and TQ effects. The degree of
nonlinearity induced by TQ, LQ, and their combination was
systematically probed in a grid of surface flux transport (SFT)
models.  The relative importance of LQ versus TQ has been found to
correlate with the ratio $u_0/\eta$ in the SFT model grid, where $u_0$
is the meridional flow amplitude and $\eta$ is diffusivity. An
analytical interpretation of this result has been given, further
showing that the main underlying parameter is the dynamo effectivity
range $\lambda_R$, which, in turn, is determined by the ratio of
equatorial flow divergence to diffusivity. The relative importance of
LQ versus TQ was found to scale as $C_1+C_2/\lambda_R^2$. 

As displayed in Fig.~\ref{fig:lambdar_quad_sep}, for various dynamo
and SFT models considered in the literature the contribution of LQ to
the amplitude modulation covers a broad range from being insignificant
to being the dominant form of feedback. On the other hand, the
contribution of a TQ effect (with the usually assumed amplitude) is
never negligible.

The role of TQ and LQ was also explored in the 2x2D dynamo model
optimized to reproduce the statistical behaviour of real solar cycles.
We demonstrated the presence of latitude quenching in the 2x2D dynamo.
The locus of the SFT component of the model in
Fig.~\ref{fig:lambdar_quad_sep} suggests that LQ contributes to the
nonlinear modulation by an amount comparable to TQ. This LQ effect is
present in the model despite the lack of any modulation in the
meridional flow.  We note that the meridional inflow module of the 2x2D model
developed by \citen{Nagy+:inflow} was not used for these runs. This
agrees with the results of \cite{Karak2020} who showed the presence of
latitude quenching in the STABLE dynamo model with a steady meridional
flow. As LQ is rather well constrained on the observational side
(\citen{Jiang:nonlin}), its presence and character may  potentially be used in the future as a test or merit function in the process of fine-tuning
of dynamo models to solar observations.

\begin{acknowledgements}
This research was supported by the Hungarian National Research,
Development and Innovation Fund (grants no. NKFI K-128384 and TKP2021-NKTA-64), by the
European Union's Horizon 2020 research and innovation programme under
grant agreement No.~955620, and by the Fonds de Recherche du Qu\'ebec
-- Nature et Technologie (Programme de recherche coll\'egiale). The
collaboration of the authors was facilitated by support from the
International Space Science Institute in ISSI Team 474.
\end{acknowledgements}

\bibliographystyle{aa}
\bibliography{SFTnonlin}

\end{document}